\documentclass[aps,prl,reprint,twocolumn,showpacs,floatfix,amsmath,amssymb,superscriptaddress,showkeys,groupedaddress]{revtex4-1}
\usepackage{graphicx}
\usepackage{epstopdf}
\usepackage{color}
\usepackage{hyperref}
\usepackage{bm}
\usepackage{printlen}
\usepackage{units}
\usepackage{bbm,pifont}
\usepackage{tikz}

\renewcommand{\vec}[1]{\mathbf{#1}}

\renewcommand{\Re}{\text{Re}}
\renewcommand{\Im}{\text{Im}}

\begin{document}
\title{WKB estimate of bilayer graphene's magic twist angles}

\author{Yafei Ren}    
\affiliation{Department of Physics, The University of Texas at Austin, Austin, Texas 78712, USA}

\author{Qiang Gao}    
\affiliation{Department of Physics, The University of Texas at Austin, Austin, Texas 78712, USA}

\author{A. H. MacDonald}    
\affiliation{Department of Physics, The University of Texas at Austin, Austin, Texas 78712, USA}

\author{Qian Niu}    
\affiliation{Department of Physics, The University of Texas at Austin, Austin, Texas 78712, USA}

\date{\today}

\begin{abstract}
    Graphene bilayers exhibit zero-energy flat bands at a discrete 
    series of magic twist angles. In the absence of intra-sublattice inter-layer
    hopping, zero-energy states satisfy a Dirac equation with a   
    non-abelian SU(2) gauge potential that cannot be diagonalized globally. 
    We develop a semiclassical WKB approximation scheme for this Dirac equation
    by introducing a dimensionless Planck's constant proportional to the twist angle, 
    solving the linearized Dirac equation around AB and BA turning points,
    and connecting Airy function solutions via bulk WKB wavefunctions.
    We find zero energy solutions at a discrete set of values of the dimensionless Planck's constant, which we obtain analytically.  Our analytic flat band twist angles correspond 
    closely to those determined numerically in previous work.
\end{abstract}

\maketitle

\textit{Introduction---.} At a discrete set of magic twist angles, bilayer graphene develops 
low-energy flat bands~\cite{Bistritzer_MacD_11, TBLG_Chiral_19} that lead to strong correlation physics including surprising 
superconductivity~\cite{TBLG1, TBLG2, TBLG3, Rev_TBLG_AMacD, TBLG_SC_Young_20, TBLG_SC_Wu_18}, novel orbital magnetism~\cite{TBLG_OrbitalMag_Law_20,  TBLG_ImagingOrbitalFM_Young_20, TBLG1_ImagingOrbitalFM_Young_20}, 
and the quantum anomalous Hall effect~\cite{TBLG1901_AHE0, TBLG1903_AHE1, TBLG1907_QAHE2}. 
The presence of narrow bands has recently been attributed to a twist-angle-dependent 
non-abelian SU(2) gauge field experienced by the 
two-dimensional (2D) Dirac fermions in bilayer graphene~\cite{NonAbelian_12_Graphene, NonAbelian_16_Graphene,TBLG_19_ZLM_Dirac, TBLG_Chiral_19}. 
In the case of 2D Dirac fermions 
with a magnetic field represented by an abelian U(1) gauge field, 
it has long been known that robust 
zero-energy states appear at any magnetic field 
strength~\cite{G_Strain_NP_10, G_Strain_PRB_10, G_Strain_PRB_13, LL_InsertNonAbelian_11}, 
with degeneracy equal to the total number of flux quanta~\cite{IndexTheorem_79, IndexTheorem_01}.  
In this work we explain why   
flat bands emerge only at quantized field strengths, corresponding to almost equally spaced 
inverse twist angles~\cite{TBLG_Chiral_19}, in the SU(2) case.

Our analysis is based on a WKB-like approximation in which we define a dimensionless effective Plank's constant $\beta$
that is proportional to twist angle.  The WKB approximation breaks down near the high-symmetry 
AB and BA stacking points in each unit cell of the 
twisted bilayer moir\'e pattern.  
Linearizing the Dirac equation around these turning points leads 
to Airy function local solutions.
By connecting the Airy functions in their asymptotic regions to the 
WKB solutions, we derive a connection formula that glues the local 
solutions together to form a global wave function. 
Because of a topological obstruction, the approximate global solution is well 
defined only at discrete $\beta$ values 
whose inverses are equally spaced.  
The set of twist angles that are defined by this condition agrees closely with those 
identified numerically in the previous work~\cite{TBLG_Chiral_19}.

\textit{Model---.} We study the chiral symmetric model of twisted bilayer graphene~\cite{NonAbelian_12_Graphene, TBLG_Chiral_19},
whose Dirac Hamiltonian $H = v_D(\vec{p}+\alpha\vec{A})\cdot \bm{\sigma}$
describes 2D Dirac fermions with velocity $v_D$,
in the presence of a non-dynamical SU(2) gauge field $\vec{A}$ that 
acts on the layer degree of freedom.
The gauge potential $\vec{A}=(A_x,A_y)=(\sum_i A_{x,i}\tau_i,\sum_i A_{y,i}\tau_i)$ arises 
physically from local-stacking-dependent interlayer tunneling and is a periodic function of position 
in the twisted bilayer.  
Here $\bm{\sigma}$ and $\bm{\tau}$ are respectively sublattice and layer
Pauli matrices.  The non-abelian nature of this equation 
arises from the nonzero commutator of $A_x$ and $A_y$.
The two gauge potential components cannot be diagonalized simultaneously,
and the equation therefore cannot be reduced to its well-understood abelian counterpart.

In this Letter we focus on zero energy solutions. 
Following Ref.~\onlinecite{TBLG_Chiral_19} we
take advantage of chiral symmetry 
to simplify the the analysis by expressing the Hamiltonian as $H=\alpha 
\mathcal{D}\sigma_- + \alpha \mathcal{D}^\dagger \sigma_+$ with $\sigma_\pm=(\sigma_x \pm i \sigma_y)/2$ and $\mathcal{D}=(p_x+ip_y)/\alpha+A_x+iA_y$. 
It follows that zero energy solutions appear in pairs that are polarized on different sublattices.  
One of the solutions satisfies $\mathcal{D}\psi= 0$,
or more explicitly upon inserting the explicit form of the 
interlayer tunneling Hamiltonian~\cite{TBLG_Chiral_19} 
\begin{align}
\label{DEq}
\left[  \begin{array}{cc}
-2i \beta  \bar{\partial } &   U(\vec{r}) \\ 
 U(-\vec{r}) & -2i \beta  \bar{\partial } \\ 
\end{array}\right]
\left[  \begin{array}{c}
\psi_{1} \\ 
\psi_{2} \\ 
\end{array}\right] = 0.    
\end{align}
where the two components are amplitudes in different layers, $\beta = \hbar /\alpha$ acts as 
an effective Plank's constant,
$U(\vec{r})=\sum_{j} e^{i(j-1)\phi}e^{-i\vec{q}_{j}\vec{r}}$ is complex, and
$\phi=2\pi/3$.  The $\vec{q}_j$ ($j=1$-$3$) are equivalent
moir\'e Brillouin zone corner wavevectors whose common magnitude 
$k_\theta=2k_{\rm D}/\sin(\theta/2)$, where $k_{\rm D}$ is the graphene Brillouin-zone corner wavevector magnitude,
is inversely proportional twist angle $\theta$ for small twists.
$U(\vec{r})$ has the translational periodicity of the moir\'e pattern.
Since the dimensionless coupling constant $\alpha$ in the Dirac equation 
is $\alpha=t_\perp/\hbar v_D k_\theta$ where 
$t_\perp \sim 100$ meV is the interlayer tunneling strength, $\beta$ is proportional to twist angle and vanishes
in the small twist angle semiclassical limit.  
Below we set $\hbar \to 1$ and measure momenta and positions in units of $k_{\theta}$ and $k_{\theta}^{-1}$.  

\begin{figure}
    \includegraphics[width=8 cm]{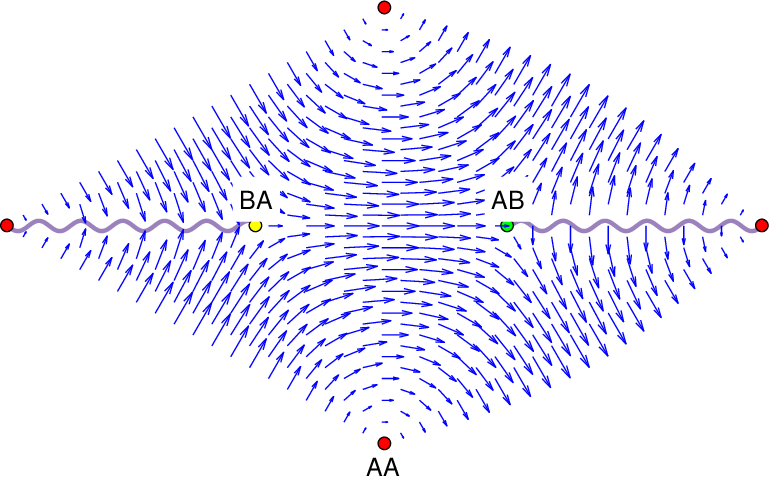}
    \caption{$A$ in the moir\'e unit cell. The real and imaginary parts of $A$ are represented 
    by the horizontal and vertical components of the plotted vectors. 
    The yellow, green, and red dots identify turning points, and the wavy lines identify the branch cut.}
    \label{FigSqrtQ}
\end{figure}

\textit{WKB approximation---.} 
Following the WKB approach we seek solutions of Eq.~\eqref{DEq}  that are linear 
combinations of $e^{iS/\beta}$ and $e^{-iS/\beta}$.  Expanding $S$ in powers of $\beta$ and truncating at leading 
order yields the two-parameter approximate form (valid for either $\psi_1$ or $\psi_2$)
\begin{align}
\label{WaveFun}
\psi= \frac{C_{+}}{\sqrt{A}} \left[  \begin{array}{c}
-\sqrt{U(\vec{r})} \\ 
\sqrt{U(-\vec{r})} \\ 
\end{array}\right] e^{iS/\beta} + 
\frac{C_{-}}{\sqrt{A}} \left[  \begin{array}{c}
\sqrt{U(\vec{r})} \\ 
\sqrt{U(\vec{-r})} \\ 
\end{array}\right] e^{-iS/\beta}.
\end{align}
where 
\begin{align}
\bar{\partial}S = A/2
\label{EqS}
\end{align}
and
$A(\vec{r})=\sqrt{U(\vec{r})U(-\vec{r})}$ is a local wavevector related to the gauge potentials $U(\pm \vec{r})$.
Eq.~\eqref{EqS} can be solved by employing a Fourier transform method that rewrites 
$A = \sum_{\vec{G}} A_{\vec{G}} e^{i\vec{G} \cdot \vec{r}} = \sum_{\vec{G}} A_{\vec{G}} e^{i\frac{1}{2} (\bar{G}z + G\bar{z})} $
where $A_{\vec{G}}$ is a Fourier coefficient, $\vec{G}=(G_x, G_y)$ is a moir\'e reciprocal latteice vector,
$G=G_x+i G_y$, and $\bar{G}=G_x-i G_y$.
Using the Fourier expansion for $A$ in Eq.~\eqref{EqS} and integrating yields 
\begin{align}
\label{Ssol}
S = S^c(z) + \frac{1}{2} A_{\vec{0}}\bar{z} +  \sum_{\vec{G}\neq \vec{0}} \frac{1}{iG} A_{\vec{G}} e^{i\frac{1}{2} (\bar{G}z + G\bar{z})} 
\end{align}
where the last term is related to the periodic spatial variation of $A$, and the 
second term to its spatial average.  The first term  $S^c(z)$ is an arbitrary function of $z$.
Requiring the imaginary part of $S$ (the log of the magnitude of the wavefunction) to be bounded
fixes $S^c(z)=\bar{A}_{\bm 0}z/2$, up to a constant.

\begin{figure}
    \includegraphics[width=9 cm]{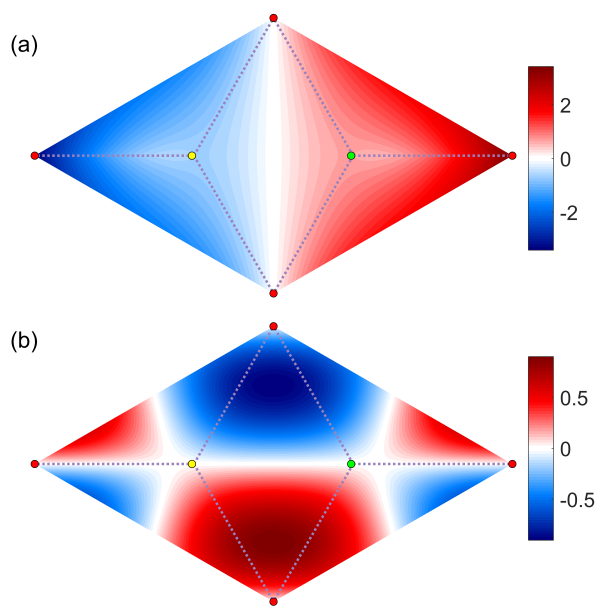}
    \caption{(a) and (b): The colormap of the real and the imaginary part of $S$. Dotted lines stand for the Stokes lines.}
    \label{FigPlotS}
\end{figure}

In Fig.~\ref{FigSqrtQ} we illustrate the dependence of $A$, which is defined only up to a sign, on position within 
one moir\'e unit cell by representing its real and imaginary parts by the horizontal and vertical components of $\vec{A}=(\Re{A}, \Im{A})$. 
This vector and its negative are the Dirac 
points of the local band structure calculated
at a given position $\vec{r}_0$ by setting $A(\vec{r})$ to its value at $\vec{r}=\vec{r}_0$.
 Eq.~\eqref{EqS} states that the complex derivative of $S$ with respect to $\bar{z}$ 
 is equal to the complex momentum $A_x+iA_y$ at each position.
 In Figs.~\ref{FigPlotS}(a) and \ref{FigPlotS}(b), the corresponding real and imaginary parts of $S$ are plotted separately. 
Importantly both real and imaginary parts of $S$ are nonzero, unlike the one-dimensional WKB case 
in which $S$ is imaginary in the classically forbidden regions and real in the classically allowed region.
By performing the derivative with respect to $z$ on both sides of Eq.~\eqref{EqS}, we find that in the 
present case $\nabla^2 \Im S = (\partial_x \Im A - \partial_y \Re A)$; the curl of the vector
$\vec{A}$ depicted in Fig.~\ref{FigSqrtQ}, which is non-zero, behaves like a source for $\Im S$:
Similarly, $\nabla^2 \Re S = \vec{\nabla} \cdot \vec{A}$, implying that 
$\vec{\nabla} \Re S$ is equal to the local wavevector up to divergence-free function.  

\begin{figure}
    \includegraphics[width=7 cm]{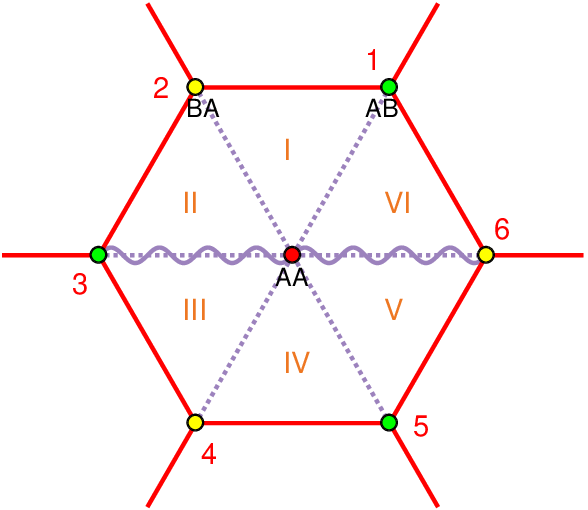}
    \caption{Stokes diagram in a moir\'e unit cell bounded by the red solid lines. The red, green, and yellow circles are turning points with local AA, AB, and BA stacking order. The branch cuts and Stokes lines are plotted by purple wavy and dotted lines, respectively. 
    Turning points at the corners are labeled 1-6 while the bounded WKB solution regions are labeled by Roman numerals I-VI.}
    \label{FigConnection}
\end{figure}

\textit{Turning points and Stokes lines---.} 
$A$ has a simple zero at the AA stacking point in the moir\'e cell, and square root singularities at both AB and BA points.
The square root singularities introduce a branch cut that lies along the wavy purple 
lines in Fig.~\ref{FigSqrtQ} when we choose the real part of A to be non-negative; it is not possible to 
choose $A$ to be both smooth and single-valued.
The two branches as a set change 
continuously, however, since $A$ connects smoothly to $-A$ across these lines.
When $|A| \gg \beta$  either branch of $A$ provides an accurate local solution to Eq.~\ref{DEq}.
Three Stokes lines can be identified that radiate from AA to AB and BA stacking points as indicated by the dotted lines in Fig.~\ref{FigPlotS} and the
purple dotted lines in Fig.~\ref{FigConnection}. 
We define the Stokes lines by the condition that 
$\vec{A} \cdot d\vec{l} = \Re A {\rm d} \bar{z}=0$,
following the definition used in the 
well studied case \cite{WKB,WKB1} where $A$ and $S$ are analytical 
functions that depend only on $\bar{z}$.  
The Stokes lines form a network that divides the two-dimensional plane into domains distinguished in Fig.~\ref{FigConnection} by Roman numerals I-VI. In the analytic case~\cite{WKB,WKB1}
the WKB solution coefficients $C_{\pm}$ change  
across the Stokes lines because the real part of $S$ is constant 
and the wavefunction stops oscillating.  Since $\Re A$ is nearly constant along the
Stokes line in our case ( see Fig.~\ref{FigPlotS}(a) ) we will assume that
we can also allow different WKB coefficients in regions that are separated by Stokes lines.
The WKB solutions are piecewise well-defined inside each domain away from turning points and Stokes lines.
To connect the WKB solutions in different domains, we need to obtain the local wavefunctions near the turning points that bridge the isolated domains. 

Near the AB and BA turning points,  $A$ vanishes, the WKB wavefunctions in Eq.~\eqref{WaveFun}
are singular, and the WKB approximation is invalid. 
A good approximation to the wavefunctions near the turning points which 
is free from singularities can be obtained by linearizing the gauge potential.  Near the 
AB points for example, $U(\vec{r}) \simeq -\frac{3}{2} e^{i\phi}(\bar{z}-\bar{z}_+)$, 
where $\bar{z}_+$ is the 
value of the $\bar{z}$ coordinate at the AB point and
$U(-\vec{r}) \simeq 3 e^{-i\phi}$.
Note that the linearized gauge potential depends only on the complex 
coordinate $\bar{z}$.
Substituting the expansion in Eq.~\eqref{DEq}, we obtain the general solution
\begin{align}
\label{WavFunAB}
{\psi}_{2} &= C_A Ai(\bar{z}') + C_B Bi(\bar{z}')  \\
{\psi}_{1} &= i2\bar{\partial}{\psi}_{2}/3e^{-i\phi} \nonumber
\end{align}
where $\bar{z}'=c(\bar{z}-\bar{z}_+)$ with $c=[9/(8 \beta^2)]^{1/3}$, 
and $C_{A,B}$ are the undetermined coefficients for Airy functions $Ai$ and $Bi$. 
It is noteworthy that both Airy functions are retained here, in constrast to the 
one-dimensional WKB analysis. 
In the one-dimensional confinement potential problems, the wavefunction is required to decay exponentially as the coordinate goes to infinity. 
The asymptotic behavior of $Bi$ increases exponentially and thus is discarded. 
Here, the solutions are confined to a finite regime in the complex plane and both functions 
can contribute to the solution.

\begin{figure}
    \includegraphics[width=7 cm]{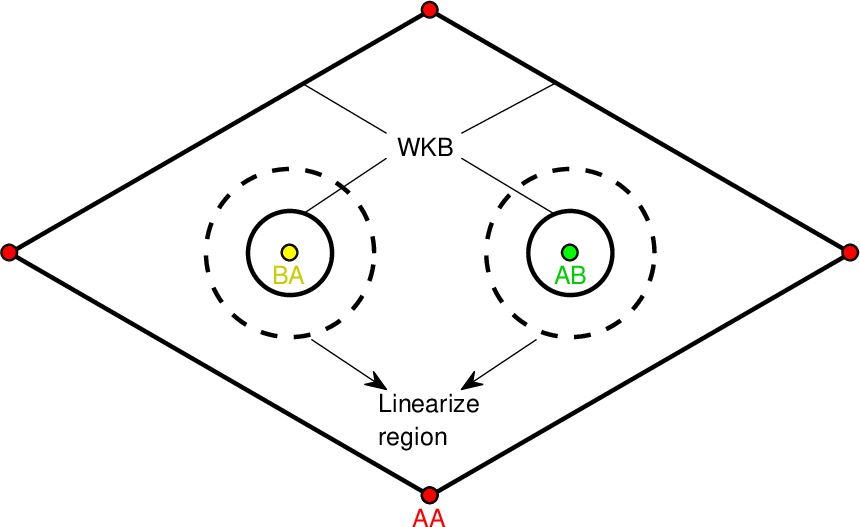}
    \caption{Validity regions of different solutions in a moir\'e unit cell. The black diamond indicates a moir\'e unit cell with red dots being AA stacking point, yellow and green dots being BA and AB stacking points. The WKB solutions are valid away from the turning points AA, AB, and BA as indicated by the regions outside the black circles in solid lines. The solutions of the linearized equation are valid inside the dashed circle.}
    \label{FigWKBregion}
\end{figure}

\textit{Connection formula---.}
Although the WKB and Airy function solutions
are obtained independently as good local approximations, 
there are regions, schematically indicated  by the area between the solid and dashed circles 
in Fig.~\ref{FigWKBregion},  where both approximations are accurate.  Inside the dashed circle centered at $\bar{z}_0$ ($1\gg |\bar{z}-\bar{z}_0|)$ , 
the linearization is justified with whereas, outside the solid circle ($|\bar{z}-\bar{z}_0| \gg (8/9)^{1/3}\beta^{2/3}$) the WKB approximation
is reliable and the Airy functions take their asymptotic form. 
One can therefore make a connection between the WKB and Airy function 
solutions by comparing them in the overlap regime.

We consider the Airy functions solved at point 1 and the WKB solutions in region I
in Fig.~\ref{FigConnection} to illustrate the connection formula derivations. 
In the region with $|\bar{z}'|\gg 1$, Eq.~\eqref{WavFunAB} can be approximated by 
\begin{align}
\psi_2 &\simeq (C_A \gamma_{A+}+C_B \gamma_{B+}) \bar{z}'^{-1/4}e^{\frac{2}{3}\bar{z}'^{3/2}} \nonumber \\
&+ (C_A \gamma_{A-}+C_B \gamma_{B-}) \bar{z}'^{-1/4}e^{-\frac{2}{3}\bar{z}'^{3/2}}
\end{align}
where $\gamma_{A,\pm}$ and $\gamma_{B,\pm}$ are asymptotic expansion coefficients for $Ai$ and $Bi$, which are different in different regions according to the argument of $\bar{z}'$ separated by the Stokes lines as detailed shown in Supplemental Materials. Specifically, for the region connecting region I to point 1 studied here, $\gamma_{A,+}=-i/2\sqrt{\pi}$, $\gamma_{A-}=1/2\sqrt{\pi}$ while $\gamma_{B,+}=1/2\sqrt{\pi}$ and $\gamma_{B-}=-i/2\sqrt{\pi}$. 

In the same asymptotic region where the WKB approximation also works, one can obtain alternatively the local solution of $S$ by substituting the linearized $U(\pm\vec{r})$ into Eq.~\eqref{EqS}. We find that around the AB point, $S(z_+,\bar{z})=i(\bar{z}-\bar{z}_+)^{3/2}/\sqrt{2} + S_{1}$ 
where the first term satisfies the differential equation 
and $S_{1}$ is a constant equal to the 
value of $S$ at turning point 1 obtained by the Fourier transform method.
By comparing the WKB and Airy solutions, we find that the exponents ($\frac{2}{3}\bar{z}'^{3/2} = (\bar{z}-\bar{z}_+)^{3/2}/\sqrt{2}\beta$) 
and prefactors ($\bar{z}'^{-1/4} = C_{\rm con} (\bar{z}-\bar{z}_+)^{-1/4}$ 
with $C_{\rm con}=c^{-1/4}$) agree.  It follows that the 
Airy functions and WKB solutions are consistent when the coefficients 
satisfy
\begin{align}
\left[  \begin{array}{c}
C_{-} \\ 
C_{+} 
\end{array}\right]_{\rm I}
=M_{\rm I,1}
\left[  \begin{array}{c}
C_{A} \\ 
C_{B} 
\end{array}\right]_{1}
\end{align}
where
\begin{align}
M_{\rm I,1} 
=\frac{1}{C_0 C_{\rm con}} &\left[  \begin{array}{cc}
e^{iS_{1}/\beta} &  \\ 
& e^{-iS_{1}/\beta} \\ 
\end{array}\right] 
\left[  \begin{array}{cc}
\gamma_{A+} & \gamma_{B+} \\ 
\gamma_{A-} & \gamma_{B-} \\ 
\end{array}\right] \nonumber,
\end{align}
and $S_1$ is the value of $S$ at point 1.

Near point 2 in Fig.~\ref{FigConnection}, which is a BA stacking point with complex coordinate $\bar{z}_-$, we find that $U(-\vec{r}) \simeq \frac{3}{2} e^{-i\phi}(\bar{z}-\bar{z}_-)$ and $U(\vec{r}) \simeq 3e^{i\phi}$. 
By redefining the variable as $\tilde{z}=-\bar{z}$ and substituting them into Eq.~\eqref{DEq}, we can obtain the local solution of $\psi_1$, instead of $\psi_2$, as a linear combination of $Ai$ and $Bi$. Identifying the  asymptotic expansion with $\psi_1$ in Eq.~\eqref{WaveFun} at point 2
also ts Airy function expansion coefficients to $C_{\pm}$.

Using their mutual relationships to $C_{\pm}$ in region I, we see that 
$C_{A,B}$ at points 2 and 1 are related
\begin{align}
\label{Tmat}
\left[  \begin{array}{c}
C_{A} \\ 
C_{B} \\ 
\end{array}\right]_1
&=
i\left[  \begin{array}{cc}
-a_{\rm 1,I} & -a_{\rm 1,R} \\ 
a_{\rm 1,R} & -a_{\rm 1,I} \\ 
\end{array}\right]
\left[  \begin{array}{c}
C_{A} \\ 
C_{B} \\ 
\end{array}\right]_2
\end{align}
where $a_1=e^{-i(S_{1}-S_{2})/\beta}$, where $S_2$ is the value of 
$S$ at point 2, and $a_{1,\rm{I/R}}$ stands for the imaginary/real parts.  

Similarly, we can obtain the connection formulas for the coefficients of Airy functions at the other neighboring turning points:
\begin{align}
T_{12}&=i\left[  \begin{array}{cc}
-a_{\rm 1,I} & -a_{\rm 1,R} \\ 
a_{\rm 1,R} & -a_{\rm 1,I} \\ 
\end{array}\right];
T_{23}= \left[  \begin{array}{cc}
\bar{a}_2 & -2ia_{\rm 2,R} \\ 
0 & -a_2 \\ 
\end{array}\right] \nonumber \\
T_{34}&=  \left[  \begin{array}{cc}
-\bar{a}_2 & -2ia_{\rm 2,R} \\ 
0 & a_2 \\ 
\end{array}\right]; \,
%
T_{45} = i\left[  \begin{array}{cc}
a_{\rm 1,I} & -a_{\rm 1,R} \\ 
a_{\rm 1,R} & a_{\rm 1,I} \\ 
\end{array}\right] \nonumber \\
T_{56}&=  \left[  \begin{array}{cc}
a_2 & -2ia_{\rm 2,R} \\ 
0 & -\bar{a}_2 \\ 
\end{array}\right]; \,\,\,\,\,
T_{61}=  \left[  \begin{array}{cc}
-a_2 & -2i a_{\rm 2,R} \\ 
0 & \bar{a}_2 \\ 
\end{array}\right]
\end{align}
where $a_2=e^{-i(S_{2}-S_{3})/\beta}$ depends
on the differences of the values of $S$ at points $2$ and $3$,
and $\bar{a}_2$ is its complex conjugate.   The phases between other neighboring turning points are the same as either $a_{1,2}$ or the conjugate of $a_{1,2}$, as guaranteed by the mirror symmetry of $A$ in our convention. This symmetry also means that the differences $S_2-S_3$ and $S_1-S_2$ are real as detailed below.
By setting the cut line at $y=0$, we find that the mirror symmetry indicates that the real and imaginary parts of $A$, i.e., $\Re A$ and $\Im A$, are separately even and odd function of $y$. Thus,  $(\partial_x \Im A - \partial_y \Re A)/2 = \Im(\partial A)$ is odd about $y=0$, which suggests that $\Im (\partial\bar{\partial} S)=\frac{1}{4}\nabla^2 \Im S$ is also odd following Eq.~\eqref{EqS}. As a result, up to a constant, $\Im S$ is an odd function about the branch cut, on which $\Im S=0$ as shown in Fig.~\ref{FigPlotS}(b). Therefore, $S_i-S_j$ is real when points $i,j$ are on the branch cut and since the amplitudes of $|a_{1,2}|=1$ equal 1
one, the determinants of all the matrices above equal 1. 

\begin{figure}
    \includegraphics[width=8 cm]{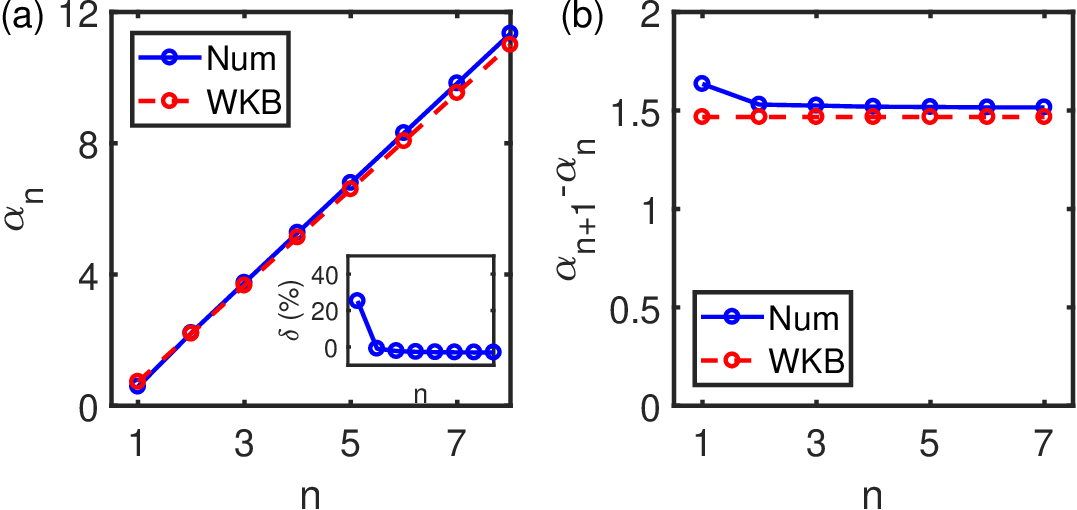}
    \caption{Comparison between numerical and WKB estimations of the magic angles via the parameter $\alpha$. Inset shows the relative error $\delta = (\alpha_n^{\rm WKB}-\alpha_n^{\rm Num})/\alpha_n^{\rm Num}$. The numerical $\alpha$ values 
    indicated by blue circles and connected with a solid line were extracted 
    from Ref.~\cite{TBLG_Chiral_19}.}
    \label{FigCompare}
\end{figure}

\textit{Quantization condition---.} 
The connection formula explained above can glue local solutions together forming a global approximation, which however has a topological obstruction. Specifically, by multiplying the connection formula $T_{12}T_{23}T_{34}T_{45}T_{56}T_{61}$, one can relate the $C_{A,B}$ at point 1 to itself. Single-valuedness of the global wavefunction requires that the product of these six matrices have at least one unit eigenvalue.  The determinant
condition guarantees that unit eigenvalues occur in pairs.  Similarly, one can also connect the coefficients at point 2 to itself through $T_{23}T_{34}T_{45}T_{56}T_{61}T_{12}$ and the same condition is required. 
The two conditions are not satisfied simultaneously  for arbitrary $\beta$,
making the global wavefunction ill-defined. The obstruction disappears at $\beta$ 
values for which $a_{\rm 2,R} = 0$, i.e., $a_2$ is purely imaginary and both matrix products reduce to the identity matrix.
The vanishing of $a_{\rm 2,R}$ imposes a constraint on the value of 
$\beta$ with $-(S_0^2-S_0^3)/\beta = -(n+1/2)\pi$, and thus $\alpha = 1/\beta = (n+1/2) \pi/(S_2-S_3) \simeq (n+1/2)1.47$.  This condition is independent of the choice of the arbitrary constant term in $S^c(z)$ since
it depends only on the difference of $S$ at two different points.
We compare this quantization condition with numerically determined
magic angles in Fig.~\ref{FigCompare}(a). The relative error 
$\delta = (\alpha_n^{\rm WKB}-\alpha_n^{\rm Num})/\alpha_n^{\rm Num}$ is shown in the inset and 
becomes very small in small twist-angle limit. The magic-angle difference ~\cite{TBLG_Chiral_19} $\alpha_{n+1}-\alpha_n$ is plotted in Fig.~\ref{FigCompare}(b) and approaches the WKB result very closely at large $n$. 

At a given $\beta$, the wavefunction of zero-energy state can be obtained by gluing the local solutions together via the connection formula. Specifically, by considering only $e^{iS/\beta}$ component in region I, we find that the coefficients $C_{A,B}$ at point 3 and point 1 are equal up to a phase factor $e^{iA_{\vec{0}}(x_3-x_1)/\beta}$ with $x_{i}$ being the $x$-coordinate of point $i$. This indicates that the wavefunctions in the regions around points 1 and 3 are identical apart from the universal phase factor. Similarly, we find that the wavefunction around point 5 is the same as that around point 1 without phase difference since the connection matrix $T_{56}T_{61}$ is an identity matrix. By repeating this procedure, one can obtain a wavefunction over the two-dimensional plane, which is invariant under translation by a moir\'e lattice vector 
up to a phase. 
Such a phase contributes to a Bloch wavevector $(A_{\vec{0}}/\beta,0)$.
A similar Bloch state can be obtained by considering only $e^{-iS/\beta}$ in region 1 with the coefficients contributing a wavevector $(-A_{\vec{0}}/\beta,0)$. Both total wavevectors show vanishing components along $y$ and are therefore distinct from the zero-energy states at the moir\'e Brillouin-zone corners $K/K'$, which occur at all twist angles and have nonzero $k_y$.

We have identified additional values of $\beta$ at which the matrix products have 
unit eigenvalues and are tridiagonal with a nonzero off-diagonal matrix element. 
In this case, only $C_A$ can be nonzero at both AB and BA stacking points,
in conflict with the  $T_{12}$ connection formula $T_{12}$ which requires a 
$C_B$ at point 1 if nonzero $C_A$ appears at point 2. Therefore, these solutions are discarded.

\textit{Summary and discussion---.}
In summary, we have studied
the chiral symmetric model of twisted bilayer graphene, which can be viewed as describing
a Dirac fermion in a non-dynamic nonabelian gauge field. 
By parameterizing the twist angle as a dimensionless effective Planck constant $\beta$, we provide a WKB solution to the zero-energy wave equation valid 
in the semiclassical limit where the gauge potential is non-zero. 
The zero-gauge-potential turning points, where the WKB approximation ceases to be valid, are located at the high symmetry moir\'e pattern points with local AB/BA/AA stacking.
We obtain a global wavefunction by linearizing the gauge potential at the AB and 
BA points, expanding the local wavefunction in terms of Airy functions, 
and matching Airy functions with WKB solutions 
in their overlapping validity regions.
For a general $\beta$, these global wavefunctions are multi-valued, exhibiting a topological obstruction. The single-valued property of a well-defined wavefunction leads to a constraint on the effective Planck constant $\beta=1/\alpha$ so that physical
zero-energy solutions are obtained at a discrete set of equally spaced $\alpha$. 
We find that these conditions are very close to the numerical values of $\alpha$ 
at which flat bands occur in numerical calculations.

By rotating the branch cuts by $\pm 2\pi/3$, additional zero energy states appear,
also at wavevectors away from the moir\'e band Dirac points which are known to have zero-energy eigenvalues at all twist angles.  Because we have only
found a finite number of zero-energy states,
our work establishes only a necessary condition for the presence of zero-energy flat bands. The semi-classical interpretation of the sufficient condition for flat bands deserves further investigation. 

We comment that the limited number of zero-energy solutions 
might be rooted in the assumption of a non-singular $S$. By relaxing the boundedness
condition on $S$ used to fix the free analytical function $S^c(z)$,
an additional periodic part can be added besides the $\bar{A}_{\vec{0}}z$ part. The resulting
wavefunction can still be well-defined if the unique wavefunction obtained above has zero amplitudes at the singular points and cancel the $S^c(z)$ singularity, following the idea from Tarnopolsky \textit{et al.}~\cite{TBLG_Chiral_19}. However, our WKB solutions have difficulty in locating zero-points since they are approximate solution valid only  
to leading order in $\beta$. 
Nevertheless, our analysis can be regarded as an alternative support to the inevitability of the singularity in the wavefunctions of zero-energy flat bands.

\textit{Acknowledgements---.}
This work was supported by DOE (DE-FG03-02ER45958, Division of Materials Science and Engineering).

\widetext
\section{Supplemental Material} 
\begin{figure*}
    \includegraphics[width=12 cm]{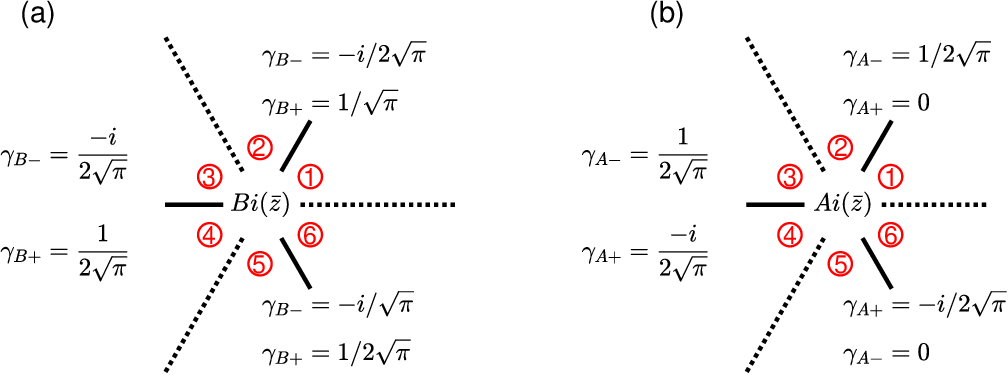}
    \caption{Asymptotic expansion coefficients in different regions. These coefficients are for variable $\bar{z}$ with the argument of it's conjugate $\arg{z} \in [0, 2\pi)$. Dotted lines are Stokes lines.}
    \label{AiBiAsympt}
\end{figure*}

In this part, we describe the derivation of the Hamiltonian $H = (\vec{p}+\alpha\vec{A})\cdot \bm{\sigma} $ and show the wavevector of our approximate WKB wavefunction. We also present the asymptotic expansion coefficients here as shown in Fig.~\ref{AiBiAsympt}.

In the absence of intra-sublattice hopping, the Hamiltonian exhibits chiral symmetry. Under the basis functions of $\{|b, A\rangle, |t, A\rangle, |b, B\rangle, |t, B\rangle \}$ where $b/t$ indicate the top or bottom layer and $A/B$ is the sublattice index, the Hamiltonian reads
\begin{align}
\mathcal{H}_0= \left[  \begin{array}{cc}
0& \mathcal{D}_0^\dagger(\vec{r}) \\ 
\mathcal{D}_0(\vec{r}) &0 \\ 
\end{array}\right]\,, \; 
\end{align}
and
\begin{align}
\mathcal{D}_0(\vec{r}) =  \hbar v_F k_\theta \left[ \begin{array}{cc}
(-2i \bar{\partial } - \kappa_b)/k_\theta & \alpha U_0(\vec{r}) \\ 
\alpha  U_0(-\vec{r}) & (-2i \bar{\partial} - \kappa_t)/k_\theta \\ 
\end{array}\right]\,
\end{align} 
where $v_F$ is the Fermi velocity of graphene,  $k_{\theta}=2k_{D}\sin(\theta/2)$ with $k_{D}=4\pi/(3a_{0})$ being the distance of Dirac point from the center of graphene's Brillouin zone and $a_0$ being lattice constant of graphene, $\bar{\partial}=(\partial_x+i\partial_y)/2$ represents the partial differential of parameter $\bar{z}=x-iy$, $\kappa_l=\kappa_{l,x}+i \kappa_{l,y}$ with the layer index being $l=t/b$ and $\bm{\kappa}_{l}=(\kappa_{l,x}, \kappa_{l,y})$ being the corresponding corners of the moir\'e Brillouin zone~\cite{Bistritzer_MacD_11, TBLG_SC_Wu_18}. $\alpha=t_\perp/\hbar v_F k_\theta$ is the inter-layer hopping strength and the interlayer hopping between different sublattices reads
\begin{align} 
U_0(\vec{r})=1+e^{i\phi}e^{-i(\vec{q}_{2}-\vec{q}_{1})\vec{r}}+e^{-i\phi}e^{-i(\vec{q}_{3}-\vec{q}_{1})\vec{r}}
\end{align}
where $\phi=2\pi/3$, moir\'e modulation vector $\vec{q}_i=k_\theta(\cos(-7\pi/6+i \phi),\sin(-7\pi/6+i \phi))$ with $i=1$-$3$. In the following, we take the units of length, wavevector, and energy as $1/k_\theta$, $k_\theta$, and $\hbar v_F k_\theta$, respectively, the Hamiltonian can be simplified with only one parameter $\alpha$~\cite{TBLG_Chiral_19}. 

To further simplify, the constant terms in the diagonal matrix elements can be removed by performing the gauge transformation that changes the reference of the momentum of bottom and top layers~\cite{TBLG_Chiral_19}
\begin{align}
\label{GaugeTransform}
\mathcal{D}_1 = \mathcal{U}^\dagger \mathcal{D}_0  \mathcal{U} =  \left[  \begin{array}{cc}
-2i \bar{\partial } & \alpha U(\vec{r}) \\ 
\alpha U(-\vec{r}) & -2i \bar{\partial } \\ 
\end{array}\right]\,
\end{align}
where 
\begin{align}
\mathcal{U}=\left[\begin{array}{cc}
e^{i\bm{\kappa}_{b}\vec{r}} &  \\ 
& e^{i\bm{\kappa}_{t}\vec{r}} \\ 
\end{array}\right],
\end{align}
and $U(\vec{r}) = e^{i(\bm{\kappa}_{t}-\bm{\kappa}_{b})\vec{r}} U_0(\vec{r}) = e^{-i\vec{q}_{1}\vec{r}} U_0(\vec{r})$.
By solving the simplified single-parameter Hamiltonian, one can obtain a wavefunction which, however, is not a Bloch wavefunction of the moir\'e lattice as the translation symmetry of $\mathcal{D}_1$ is different from the moir\'e lattice and one need to perform the inverse gauge transform shown in Eq.~\eqref{GaugeTransform}, which will leads to a plane wave part with wavevector $\vec{k}_\nu=\vec{q}_3+\vec{q}_1/2$. 

\end{document}